\documentclass{IEEEoj-data}
\usepackage{cite}
\usepackage{amsmath,amssymb,amsfonts}
\usepackage{algorithmic}
\usepackage{graphicx,xcolor}
\usepackage{textcomp}
\usepackage{hyperref}
\usepackage{balance}
\definecolor{main}{HTML}{CFCFCF}
\definecolor{sub}{HTML}{CFCFCF}
\usepackage{flushend}
\usepackage{multirow}
\usepackage{fontawesome}
\usepackage{listings}

\definecolor{darkgreen}{rgb}{0,0.5,0}
\lstdefinestyle{cppStyle}{
  language=C++,
  basicstyle=\ttfamily\small,
  commentstyle=\color{darkgreen}\textbf,
  showstringspaces=false
}
\begin{document}
% \receiveddate{\today}
% \reviseddate{11 April, 2024}
% \accepteddate{00 April, 2024}
% \publisheddate{00 May, 2024}
% \currentdate{24 June, 2024}
% \doiinfo{DD.2024.0624000}
\let\WriteBookmarks\relax
\def\floatpagepagefraction{1}
\def\textpagefraction{.001}

%\shorttitle{CppSATD: A Reusable Self-Admitted Technical Debt Dataset in C++}    

%\shortauthors{Pham et al.}  

\title{\textcolor{black}{Descriptor:} \textcolor{ieeedata}{C++ Self-Admitted Technical Debt Dataset (CppSATD)}}  


% Title footnote text
%\tnotetext[1]{This work was supported by ... (add funding note if applicable)}

\author{Phuoc Pham\authorrefmark{1}, Murali Sridharan\authorrefmark{1}, Matteo Esposito\authorrefmark{1}, and Valentina Lenarduzzi\authorrefmark{1}}
\affil{University of Oulu, Oulu, Finland}
\corresp{CORRESPONDING AUTHOR: Matteo Esposito (e-mail: matteo.esposito@oulu.fi).}
\authornote{Phuoc Pham and Murali Sridharan contributed equally to this article. }
\markboth{Preparation of Papers for IEEE DATA DESCRIPTIONS}{Pham \textit{et al.}}
\begin{abstract}
In software development, technical debt (TD) refers to suboptimal implementation choices made by the developers to meet urgent deadlines and limited resources, posing challenges for future maintenance. Self-Admitted Technical Debt (SATD) is a sub-type of TD, representing specific TD instances ``openly admitted'' by the developers and often expressed through source code comments. Previous research on SATD has focused predominantly on the Java programming language, revealing a significant gap in cross-language SATD. Such a narrow focus limits the generalizability of existing findings as well as SATD detection techniques across multiple programming languages. Our work addresses such limitation by introducing CppSATD, a dedicated C++ SATD dataset, comprising over 531,000 annotated comments and their source code contexts. Our dataset can serve as a foundation for future studies that aim to develop SATD detection methods in C++, generalize the existing findings to other languages, or contribute novel insights to cross-language SATD research.
\\ 

 {\textcolor{ieeedata}{\abstractheadfont\bfseries{IEEE SOCIETY/COUNCIL}}}     Computer Society (CS)\\  
 \\
 {\textcolor{ieeedata}{\abstractheadfont\bfseries{DATA DOI/PID}}} \href{https://doi.org/10.5281/zenodo.15275192}{10.5281/zenodo.15275192}\\ 
  
 {\textcolor{ieeedata}{\abstractheadfont\bfseries{DATA TYPE/LOCATION}}} C++ source code and code comments; Oulu, Finland
\end{abstract}


\maketitle

\section*{BACKGROUND}
\label{sec:background}

Technical debt (TD) manifests when developers have to choose quick, convenient solutions instead of more thorough and optimal ones \cite{cunningham1992wycash}. Development teams usually work under pressure due to user demands, with limited resources and time-to-market \cite{esposito_early_2023}. Hence, practitioners might intentionally or unintentionally accumulate TD, which can lead to increased maintenance costs, poor system performance, and an overall degradation in code quality \cite{esposito_can_2023}. Self-Admitted Technical Debt (SATD) is a sub-type of TD that represents SATD instances purposely admitted by the developers for future improvements, and typically reported via source code comments, commit messages, pull requests, or issue records. Based on the described characteristics of the issues according to the practitioners and their locations, we can categorize SATD into various types \cite{alves_ontology_technical_debt}. Specifically, \cite{alves_ontology_technical_debt} proposed 13 SATD types differentiated by their characteristics in software systems, including architecture, build, code, defect, design, documentation, infrastructure, people, process, requirement, service, test automation, and test debt.

\cite{da2017using} identified SATD from over 62,000 comments of 10 Java projects and classified them into five SATD types (i.e., design, requirement, defect, test, and documentation), which are the only SATD types that can commonly be found in code comments \cite{alves_ontology_technical_debt}. Such dataset has been widely employed in the literature \cite{sheikhaei2024empirical,santos_lstm,chen_xgboost,yu_gan}, and is commonly referred to as Maldonado-62k dataset. However, current research efforts dominantly focus on Java, revealing a gap of cross-language datasets. Moreover, the Maldonado-62k dataset does not include surrounding code contexts, which is an important component that might improve the overall accuracy of detection models. Thus, we introduce CppSATD, which is the first C++ SATD dataset with multiclass SATD annotations and surrounding code contexts, thereby allowing researchers to broaden their scope to another popular programming language and open a new avenue of research for SATD types.

In this study, we combine \textbf{design} and \textbf{code debt} as \textbf{one category} due to their high similarity in both characteristics and definitions, resulting in five main SATD categories: Design/Code, Requirement, Defect, Test, and Documentation. Table \ref{tab:satd_types_definitions} presents the definitions and indicators of each SATD type, according to \cite{alves_ontology_technical_debt} and \cite{farias_code_comment_analysis}.

\begin{table*}[t]
\centering
\scriptsize 
\caption{Definitions of SATD types. Adapted from \cite{alves_ontology_technical_debt} and \cite{farias_code_comment_analysis}.}
\label{tab:satd_types_definitions}
\resizebox{\linewidth}{!}{
\begin{tabular}{p{2cm}|p{7cm}|p{5cm}}
\hline
\multicolumn{1}{l|}{{\textbf{\begin{tabular}[c]{@{}c@{}}SATD Type\end{tabular}}}} &
\multicolumn{1}{c|}{{\textbf{\begin{tabular}[c]{@{}c@{}}Definition \\ \end{tabular}}}} &
\multicolumn{1}{c}{\textbf{\begin{tabular}[c]{@{}c@{}}Indicator \\ \end{tabular}}}
\\ \hline
\vspace{1pt} Design & Refers to debt that can be discovered by analyzing the source code by identifying the use of practices which violated the principles of good object-oriented design (e.g. very large or tightly coupled classes).  & 

- Violations of the principles of good object-oriented design;

- Some types of code smell; 

- Complex classes or methods;

- Excessive design complexity. \\

\vspace{9pt} Code & Refers to the problems found in the source code which can affect negatively the legibility of the code making it more difficult to be maintained. Usually, this debt can be identified by examining the source code of the project considering issues related to bad coding practices.  &   

- Unnecessary code duplication and complexity;

- Bad style that reduces the readability of code;

- Over-complex code. \\

\vspace{18pt} Requirement & Requirements debt refers to tradeoffs made with respect to what requirements the development team need to implement or how to implement them. Some examples of this type of debt are: requirements that are only partially implemented, requirements that are implemented but not for all cases, requirements that are implemented but in a way that doesn’t fully satisfy all the non-functional requirements (e.g. security, performance, etc.).  &   

- Requirements that are only partially implemented. \\

\vspace{22pt} Defect & Software projects may have known and unknown defects in the source code. Defect debt consists of known defects, usually identified by testing activities or by the user and reported on bug track systems, that the development team agrees should be fixed but, due to competing priorities and limited resources, have to be deferred to a later time. Decisions made by the development team to defer addressing defects can accumulate a significant amount of TD for a product making it harder to fix them later.  &   

- Postponed decisions on fix defects, bugs, or failures found in software systems. \\

\vspace{5pt} Test & Refers to issues found in testing activities which can affect the quality of testing activities. Examples of this type of debt are planned tests that were not run, or known deficiencies in the test suite (e.g. low code coverage). &   

- Insufficient test coverage;

- Lack of tests (e.g., unit tests, integration tests, and acceptance tests); 

- Deferred testing;

- Lack test case planning. \\

\vspace{10pt} Documentation & Refers to the problems found in software project documentation and can be identified by looking for missing, inadequate, or incomplete documentation of any type. Inadequate documentation is those that currently work correctly in the system, but fail to meet certain quality criteria of software projects.  &   

- Missing documentation;

- Inadequate documentation; 

- Outdated documentation;

- Incomplete documentation. \\
\hline
\end{tabular}
}
\end{table*}

\section*{COLLECTION METHODS AND DESIGN}
\label{sec:methods}

\begin{figure*}[t]
\centering
\includegraphics[width=\textwidth]{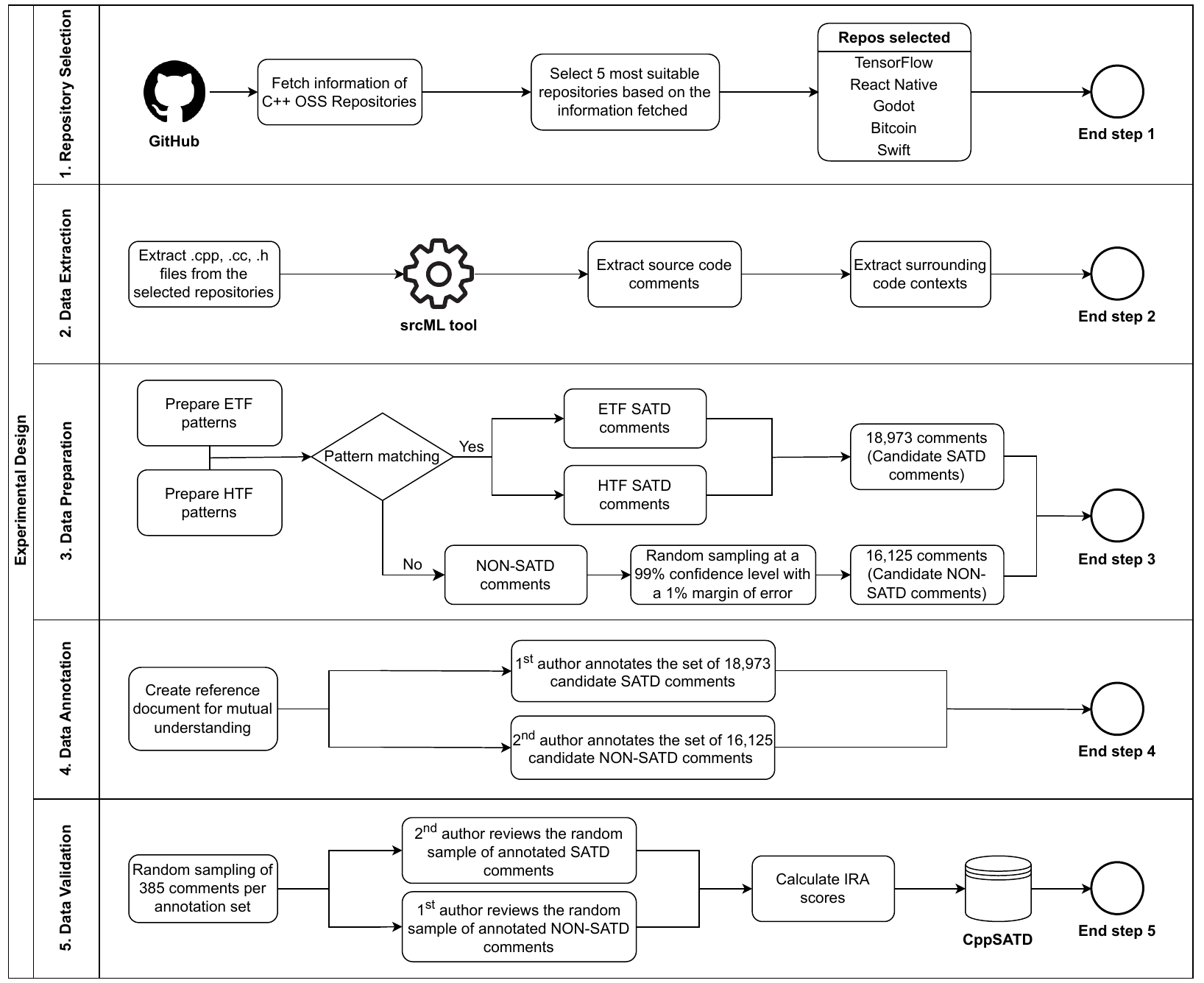}
\caption{An overview of experimental design.}
\label{fig:study_design}
\end{figure*}

In this section, we present our methodology, outlining the systematic process we followed to construct the CppSATD dataset. We design our study workflow in line with the Business Process Model and Notation (BPMN) 2.0 \cite{model2011notation} (Figure \ref{fig:study_design}), and divided it into five main steps: (1) repository selection, (2) data extraction, (3) data preparation, (4) data annotation, and (5) data validation. We cover the details of the first four steps in this section, while the next section describes further how we performed the last step of data validation. For a quick overview, according to Figure \ref{fig:study_design}, we selected the most suitable repositories based on three criteria (popularity, activeness, and diversity of domains), which resulted in five repositories: TensorFlow, React Native, Godot, Bitcoin, and Swift (\textbf{Step 1}). From each repository, we extracted C++ class files (.cpp and .cc files) and header files (.h files) to collect the source code comments. Using the comment locations, we retrieved their preceding and succeeding code snippets (\textbf{Step 2}). In \textbf{step 3}, we leveraged the SATD patterns, including both Easy-To-Find (ETF) and Hard-To-Find (HTF) patterns from the work of \cite{sridharan2023pentacet} to extract the comments that were most likely to be SATD for our manual annotations. The remaining comments that did not match the SATD patterns were conjectured to be NON-SATD. To assess such a conjecture, we randomly chose a sample with its size calculated to ensure a 99\% confidence level and a 1\% margin of error. This step generated two data subsets: a candidate SATD set (likely SATD comments) and a candidate NON-SATD set (comments with no evident SATD presence). In \textbf{step 4}, the authors manually annotated both subsets: the first author worked on the candidate SATD comments, while the second author was responsible for the candidate NON-SATD comments. Finally, we validated the correctness of our manual annotations by measuring the Inter-Rater Agreement (IRA) score to ensure that the annotations were consistent and reliable without any bias (\textbf{Step 5}). The details of each step are described in the following sections.

\subsection{Repository Selection}
\label{repo_selection}

In step 1 (Figure \ref{fig:study_design}), we selected the five suitable C++ OSS repositories based on three criteria: popularity, activeness, and the diversity of application domains. These requirements ensure that the chosen repositories precisely represent the characteristics of C++ programming language and that our study can contribute to a broad span of audience. To evaluate whether a repository could meet the established criteria, we fetched different information about the repositories from GitHub, including the total number of stars, the total number of commits, the number of commits since 2022, and the application domains. The number of stars can reflect the popularity \cite{rubaye_popularity}, while consistent recent commit history is an indicator of how active a project is. Moreover, we also made sure that the selected repositories could cover different domains of applications. By checking all respective information, we were able to identify five appropriate C++ repositories for our study, which are TensorFlow, React Native, Godot, Bitcoin, and Swift. Specifically, the five chosen repositories belong to five different domains, having 65,000 stars at a minimum and maintaining a significant number of commits since 2022 (Table \ref{tab:repo_metadata}), according to the data collection date (May 7, 2024).

\begin{table}[t]
\centering
\caption{CppSATD Dataset: OSS C++ Repositories Metadata}
\label{tab:repo_metadata}
\resizebox{\linewidth}{!}{
\begin{tabular}{l|r|r|r|r|r}
\hline
\textbf{Repository} &
\textbf{Version} &
\textbf{Total Stars} &
\textbf{Total Commits} &
\textbf{Total Commits since 2022} & 
\textbf{Domain} \\ \hline
tensorflow & 2.15.1 & 182,538 & 172,409 & 43,342 & AI/ML \\
react-native & 0.74.1 & 115,953 & 39,306 & 9,431 & Software Development\\
godot & 4.1.4-stable & 84,000 & 80,303 & 29,763 & Game Development \\
bitcoin & 27.0 & 76,176 & 44,784 & 9,334 & Crypto \& Blockchain \\
swift & 5.10 & 65,973 & 195,795 & 45,603 & Programming Language\\ \hline
\end{tabular}
}
\end{table}

\subsection{Data Extraction}
\label{data_extraction}

\begin{figure*}[t]
\centering
\includegraphics[width=0.7\textwidth]{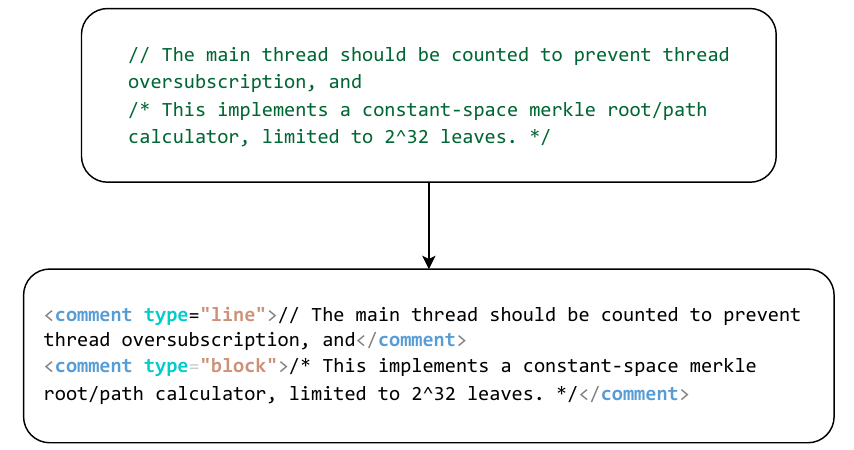}
\caption{XML format transformation.}
\label{fig:xml_format_transformation}
\end{figure*}

In step 2 (Figure \ref{fig:study_design}), we extract the source code comments and their surrounding code contexts. We focus solely on C++ class and header files, as they are the file types that constitute the source code and architecture of a C++ project. Subsequently, we utilized the srcML \cite{collard2013srcml} tool to transform the collected class and header files to the XML format, where source code and comments are wrapped within tags. Figure \ref{fig:xml_format_transformation} shows an example of how a line comment and a block comment are transformed to their XML representations. With the comments encapsulated within tags, we were able to apply simple pattern-based approach to extract comments from the XML files generated. We then used the retrieved comments to trace back to their original class and header files to collect their preceding and succeeding code contexts. In the case of an empty line before or after the comments, we treated the corresponding context as empty. We eventually managed to extract a total of \textbf{531,367 comments} along with their code contexts from the five repositories selected.

\subsection{Data Preparation}
\label{data_pre}

In step 3 (Figure \ref{fig:study_design}), since manually classifying over 531k comments would be an error-prone task, we filtered out the comments that were most likely to be SATD at our best efforts with a pattern-based filtering approach, and only manually annotated this subset, making the process more efficient. To perform this filtering step, we leveraged the Easy-to-Find (ETF) SATD patterns and Hard-to-Find (HTF) SATD patterns proposed in the work of \cite{sridharan2023pentacet}. In particular, ETF SATD refers to instances of SATD that can be easily spotted using simple keywords or patterns, such as `\texttt{todo}', `\texttt{fixme}', `\texttt{revisit}', etc. On the other hand, HTF SATD refers to more complicated features such as `\texttt{should be unnecessary}', `\texttt{need[s]* to be improved}', `\texttt{cheat here}', etc. Overall, we used 21 ETF SATD text patterns and 656 HTF text patterns to detect the most likely SATD comments. This best-effort filtering step resulted in \textbf{13,739 ETF} SATD comments and \textbf{5,234 HTF} SATD comments, adding up to a total of \textbf{18,973 candidate} SATD comments for our manual annotation (Table \ref{tab:etf_htf_table}).

To validate that the remaining set of 512,394 comments is truly NON-SATD, we took a random sample of comments, whose sampling size was calculated using Equation \ref{eq:sample_size_calculation}. The sampling size is based on the total number of the remaining comments (512,394) with a confidence level of 99\% and a 1\% margin of error, leading to a total of 16,125 comments to be manually annotated. We chose the confidence level of 99\% due to the large size of the remaining set and the fact that this dataset is targeted to be used for model training purposes, which typically require reliable training data. Table \ref{tab:random_non_satd_sample} depicts in detail the number of comments sampled for each repository. Eventually, we collected 18,973 candidate SATD comments and a sample of 16,125 candidate NON-SATD comments for the next step.

\begin{equation}
\label{eq:sample_size_calculation}
n = \frac{z^2 \cdot p \cdot (1 - p)}{E^2}
\end{equation}
where:
\begin{itemize}
  \item \( n \) is the required sample size
  \item \( z \) is the z-score corresponding to the desired confidence level (1.96 for 95\% confidence/2.58 for 99\%)
  \item \( p \) is estimation proportion, we use 0.5, which provides the maximum possible sample size, ensuring adequate representation across all SATD classes.
  \item \( E \) is the margin of error (0.05 for 5\% error rate/0.01 for 1\% error rate)
\end{itemize}

\begin{table}[t]
\centering
\caption{Candidate SATD: Number of comments extracted by ETF and HTF patterns}
\label{tab:etf_htf_table}
\resizebox{0.8\linewidth}{!}{
\begin{tabular}{@{}l|r|r|r}
\hline
\textbf{Repository Name} &
\textbf{ETF SATD comments} &
\textbf{HTF SATD comments} &
\textbf{Total} \\ \hline
tensorflow & 6,515 & 1,333 & 7,848 \\
react-native & 252 & 117 & 369 \\
godot & 2,109 & 1,617 & 3,726 \\
bitcoin & 217 & 277 & 494 \\
swift & 4,646 & 1,890 & 6,536 \\ \hline
\textbf{Total} & 13,739 & 5,234 & 18,973 \\ \hline
\end{tabular}
}
\end{table} 
\begin{table}[t]
\centering
\caption{Candidate NON-SATD: Number of Comments per Repository}
\label{tab:random_non_satd_sample}
\resizebox{\linewidth}{!}{
\begin{tabular}{@{}l|r|r}
\hline
\textbf{Repository Name} &
\textbf{Remaining Comments (After ETF \& HTF Filtering)} &
\textbf{Candidate NON-SATD Comments} \\ \hline
tensorflow & 244,402 & 6,839 \\
react-native & 7,952 & 285 \\
godot & 134,983 & 4,819 \\
bitcoin & 26,291 & 847 \\
swift & 98,766 & 3,335 \\ \hline
\textbf{Total} & 512,394 & 16,125 \\ \hline
\end{tabular}
}
\end{table} 

\subsection{Data Annotation}
\label{data_annotate}

In step 4 (Figure \ref{fig:study_design}), we manually annotate the two collections of candidate SATD and NON-SATD comments obtained from the previous step. The authors agreed on the annotation reference document (RD), which provided detailed explanations about the relevant concepts, definitions, related works, and examples of SATD types, serving as a guideline to ensure that both annotators had an aligned understanding of SATD types and preserved our consistency throughout the annotation process. Moreover, the first author also performed a mock annotation of source code comments into SATD types, including both ETF and HTF SATD, to document these as illustrative examples in the RD. After confirming that we had a uniform overview of SATD-type definitions and finalizing the RD as a guideline, we started the manual annotation procedure, with the first author annotating the set of 18,973 SATD comments and the second author handling the sample of 16,125 NON-SATD comments.

In particular, the first author manually classified 18,973 comments with their code contexts, resulting in 13,044 SATD comments and 5,929 NON-SATD comments (Table \ref{tab:satd_annotation}). Within the five SATD types, design/code and requirement debt are the two most prevalent categories, accounting for approximately 64\% of the total comments, whereas the proportion of documentation debt is remarkably lower than other SATD types, with only 46 instances (0.24\%). During the annotation process, we also encountered certain special cases where a comment could be classified into two or more types of SATD. With those scenarios, we decided to choose the category that reflected the core problem to be addressed, which was an approach followed by \cite{maldonado_2015}. For example, consider the following SATD comment with its code context, although the SATD comment points out a test inconsistency between the two environments - which can be interpreted as test debt, the underlying issue causing the inconsistency and test failure is a defect of the system, and thus, it should be classified as defect debt.

\begin{lstlisting}[style=cppStyle, breaklines=true, breakatwhitespace=false]
// TODO(alanchiao): this passes locally, but fails on continuous build system.
// Re-enable when root cause found.
TEST_P(ConvolutionOpTest, DISABLED_PointwiseMultifilterFloat32) {
  ConvolutionOpModel m(GetRegistration(), {TensorType_FLOAT32, {2, 2, 4, 2}},
                       {TensorType_FLOAT32, {2, 1, 1, 2}},
                       {TensorType_FLOAT32, {}}, 1, 1);
\end{lstlisting}

Regarding the NON-SATD random sample of 16,125 comments, the second author was able to identify 25 SATD comments from a total of 16,125 comments (Table \ref{tab:non_satd_annotation}). This means that the probability of SATD instances is 0.155\% in the collection of 512,394 NON-SATD comments. Hence, we can conclude with a confidence level of 99\% that the remaining set of 512,394 NON-SATD comments has less than 1\% chance of containing some SATD comments. Accordingly, we applied the NON-SATD label to all leftover comments after ETF and HTF pattern filtering, thereby finalizing the data annotation.

\begin{table}[t]
\caption{Candidate SATD Comments: Proportions of SATD Types}
\label{tab:satd_annotation}
\resizebox{\columnwidth}{!}{
\begin{tabular}{l|c|c|c|c|c|c|r}
\hline
\multirow{2}{*}{\textbf{Repos}} & \multicolumn{6}{c|}{\textbf{\# (\%)}} & \multirow{2}{*}{\textbf{Total}} \\ \cline{2-7}
& \textbf{Design/Code} &
  \textbf{Requirement} &
  \textbf{Defect} &
  \textbf{Test} &
  \textbf{Documentation} &
  \textbf{NON-SATD} \\ \hline
  
  tensorflow &
    3,698 (47.12) &
    1,988 (25.33) &
    208 (2.65) &
    368 (4.69) &
    29 (0.37) &
    1,557 (19.84) & 
    7,848 \\
 
  react-native &
   162 (43.90) &
   55 (14.91)	&
   7 (1.90) &
   13 (3.52) &
   1 (0.27) &
   131 (35.50) & 
   369 \\
  
  godot &
   1,290 (34.62) &
   589 (15.81) &
   101 (2.71) &
   22 (0.59)	&
   11 (0.30) &
   1,713 (45.97) & 
   3,726 \\
  
  bitcoin &
   120 (24.29) &
   42 (8.50) &
   16 (3.24) &
   9 (1.82) &
   0 (0.00) &
   307 (62.15) & 
   494 \\
  
  swift &
   2,955 (45.21) &
   1,239 (18.96) &
   106 (1.62) &
   10 (0.15) &
   5 (0.08) &
   2,221 (33.98) & 
   6,536 \\ \hline

\textbf{Total} &
   8,225 (43.35) &
   3,913 (20.62) &
   438 (2.31) &
   422 (2.22) &
   46 (0.24) &
   5,929 (31.25) & 
   18,973 \\ \hline
\end{tabular}
}
\end{table} 
\begin{table}[]
\caption{Candidate NON-SATD Comments: Proportions of SATD Types}
\label{tab:non_satd_annotation}
\resizebox{\columnwidth}{!}{
\begin{tabular}{l|c|c|c|c|c|c|r}
\hline
\multirow{2}{*}{\textbf{Repos}} & \multicolumn{6}{c|}{\textbf{\# (\%)}} & \multirow{2}{*}{\textbf{Total}} \\ \cline{2-7}
& \textbf{Design/Code} &
  \textbf{Requirement} &
  \textbf{Defect} &
  \textbf{Test} &
  \textbf{Documentation} &
  \textbf{NON-SATD} \\ \hline
  
  tensorflow &
    6 (0.09) &
    3 (0.04) &
    3 (0.04) &
    0 (0.00) &
    0 (0.00) &
    6,827 (99.82) & 
    6,839 \\
 
  react-native &
   2 (0.70) &
   0 (0.00) &
   0 (0.00) &
   0 (0.00) &
   0 (0.00) &
   283 (99.30) & 
   285 \\
  
  godot &
   3 (0.06) &
   0 (0.00) &
   1 (0.02) &
   0 (0.00)&
   0 (0.00) &
   4,815 (99.92) & 
   4,819 \\
  
  bitcoin &
   1 (0.12) &
   0 (0.00) &
   0 (0.00) &
   0 (0.00) &
   0 (0.00) &
   846 (99.88) & 
   847 \\
  
  swift &
   5 (0.15) &
   1 (0.03) &
   0 (0.00) &
   0 (0.00) &
   0 (0.00) &
   3,329 (99.82) & 
   3,335 \\ \hline

\textbf{Total} &
   17 (0.11) &
   4 (0.02) &
   4 (0.02) &
   0 (0.00) &
   0 (0.00) &
   16,100 (99.84) & 
   16,125 \\ \hline
\end{tabular}
}
\end{table} 

\section*{VALIDATION AND QUALITY}
\label{sec:validation_quality}
To evaluate the soundness of our manual annotations, we calculated the Inter-Rater Agreement (IRA) using Cohen's kappa coefficient introduced by \cite{cohen1960coefficient}. In particular, Cohen's kappa coefficient evaluates the agreement and provides us with a statistical mean that can assess whether the agreement is not observed by chance, making it more accurate than other simple percent agreement calculations. The coefficient ranges from -1 to +1, and the closer the value is to +1, the more solid the agreement, which is formulated as follows:

\begin{equation}
\label{eq:ira_formula}
k = \frac{p_o - p_c}{1 - p_c}
\end{equation}

where:
\begin{itemize}
    \item $p_o =$ the proportion of units in which the raters agree.
    \item $p_c =$ the proportion of units for which agreement is expected by chance.
\end{itemize}

To compute Cohen's kappa coefficient, we randomly selected two representative samples for the two manually annotated data subsets. We determined the sampling sizes based on a 95\% confidence level and a 5\% margin of error, following Equation \ref{eq:sample_size_calculation} and the methodology adopted by \cite{sridharan2023pentacet, falessi_enhancing_2023, esposito_extensive_2024}. This yielded sample sizes of 374 for the 13,044 SATD comments identified by the manual annotation of the first author, and 376 for the set of 16,125 candidate NON-SATD comments annotated by the second author. To promote consistency, we scaled both to 385. We stratified the SATD sample by SATD type (including both ETF and HTF SATD), and the NON-SATD sample by SATD/NON-SATD label.

Once the two representative samples were all set, the two authors independently reviewed them for the IRA calculation. Consequently, the SATD annotation sample resulted in a Cohen's kappa coefficient of 0.86, while the NON-SATD annotation sample gave a coefficient of 1.0. According to \cite{SimWright2005}, kappa scores greater than 0.80 are considered excellent (Table \ref{tab:kappa-agreement}). These high IRA scores reveal excellent consistency, validating the reliability of our manual classifications \cite{esposito_beyond_2024, esposito_leveraging_2024}.

\begin{table}[t]
\centering
\caption{Interpretation categories for agreement levels by $\kappa$ value}
\label{tab:kappa-agreement}
\resizebox{\linewidth}{!}{%
\begin{tabular}{c|c|c|c|c}
\hline
$\kappa < 0$ & $0 \leq \kappa < 0.4$ & $0.4 \leq \kappa < 0.6$ & $0.6 \leq \kappa < 0.8$ & $0.8 \leq \kappa < 1$ \\
\hline
None & Poor  & Discrete  & Good  & Excellent  \\
\hline
\end{tabular}%
}
\end{table}

\section*{RECORDS AND STORAGE}
\label{sec:records_storage}
The CppSATD dataset includes a total of 531,367 comments with their surrounding code contexts, which are annotated as one of the five SATD types or NON-SATD. The dataset contains seven fields: comment\_id (Unique comment identifier within a project), commenttext, preceding\_code (Code before the comment), succeeding\_code (Code after the comment), file\_id (Unique file identifier within a project), projectname, and annotation. We provide a replication package publicly available at Zenodo \cite{pham_2025_15562944}, comprising: 
\begin{itemize}
     \item \textbf{cppsatd.csv:} The entire CppSATD dataset.
     
     \item \textbf{manual\_annotations.csv:} Separated from the CppSATD dataset, this file includes 35,098 comments that the authors manually annotated.
     
     \item \textbf{Annotation\_Reference\_Document.pdf:} The annotation RD serving as a guideline for our manual annotations. 
     
     \item \textbf{repos.zip:} The source code of the five C++ repositories selected, from which we collected the data.
     
     \item \textbf{cppsatd.data.extraction.zip:} The Python scripts and text patterns used for the data extraction.
\end{itemize}

% \section*{Data Description}
% \label{sec:data_description}
% \input{Sections/Data_Description}

% \section*{Related Work}
% \label{sec:related_work}
% \input{Sections/Related Work}

% \section*{DISCUSSION}
% \label{sec:discussion}
% \input{Sections/Discussion}

\section*{INSIGHTS AND NOTES}
\label{sec:insights_and_notes}
Future studies can use CppSATD for purposes including, but not limited to:

\begin{itemize}
    \item \textbf{Identifying common SATD patterns:} As CppSATD includes a large number of comments classified into multiple SATD types with their code contexts, it may support future research in discovering novel SATD patterns that have remained unexplored due to the current limited data resources. Such findings can provide a more reliable collection of recurring SATD patterns that allow researchers to develop better pattern-based detection and classification techniques, enhancing the ability to prevent debt accumulation. On top of that, these patterns might offer helpful insights into the characteristics of SATD types, which further enrich our understanding of SATD in software systems.
    
    \item \textbf{Enhancing SATD detection tools:} Most of the existing SATD detection tools were developed using datasets on the Java programming language, particularly the Maldonado-62k dataset \cite{da2017using}, which limits the scope of these tools to only Java repositories. Furthermore, the available datasets often lack the incorporation of source code contexts, which makes the detection tools tied solely to the code comments, dropping important information that may improve the effectiveness of SATD detection techniques. Therefore, the introduction of CppSATD will not only broaden the research scope to more programming languages but also support the development of more robust detection and classification techniques.
    
    \item \textbf{Analyzing the impact of language-specific characteristics on SATD:} With the limited data resources, there have been no studies discovering the difference in TD levels between different programming languages. Such comparisons will provide insights into language-specific factors that may influence the frequency of SATD occurrences in software systems, such as source code structure, code syntax, error handling, or documentation culture, growing our understanding regarding how different programming languages impact TD.
    
    \item \textbf{Understanding the evolution of SATD types in C++:} Tracing the lifecycle of TD in software systems has been a crucial area that was studied by \cite{potdar2014exploratory,digkas_evolution} and \cite{sutoyo_tracing}, as it can showcase how SATD evolves in a project. Particularly, with the CppSATD dataset, the researchers will be able to examine the lifecycle of a SATD type in C++ software systems and uncover why a TD is introduced, when SATD instances are documented, how long it takes to remove them, and which type of SATD is most time-consuming to be resolved. The findings from these time-associated analyses will allow practitioners to develop corresponding strategies for each SATD type to better address the related issues, mitigate TD earlier in a software development life cycle (SDLC), and enhance long-term software maintainability.
\end{itemize}

Considering the potential of our dataset, we believe CppSATD can serve as a foundation for multiple promising extensions, including but not limited to the following areas:

\begin{itemize}
    \item Our multi-class SATD dataset will open future work in applying supervised learning models for SATD detection in C++. We can establish novel baseline benchmarks on C++ by using LLMs such as StarCode or QwenCoder fine-tuned on this dataset \cite{esposito_generative_2025}. Moreover, we can extend the dataset by including additional metadata, such as cyclomatic complexity, lines of code (LOC), and comment length, supporting further research that requires rich contextual information.

    \item By training models on the CppSATD dataset, future work can include integrating SATD detection directly into developer environments, such as creating real-time classification plugins for IDEs (e.g., VSCode, CLion). In addition, we can expose models via APIs, possibly through Hugging Face, to allow researchers and developers to easily access pre-trained SATD classifiers for integration into tools, automated analysis pipelines, or custom applications on private codebases.

    \item Economic impact is also a meaningful aspect that should be explored. Utilizing the dataset along with temporal metadata, we can analyze the cost trade-offs between different stages of SATD life cycle and estimate the return on investment (ROI) of SATD handling strategies, creating a solid bridge between academic research and industry practices.

    \item Especially, as current research landscape revolves around Java programming language, future work can include cross-language research, such as comparative analysis between C++, Python, and Java. These comparative studies can provide insights into the differences of SATD characteristics between programming languages, thereby explaining how language-specific factors can influence the occurrences of SATD in software systems.
\end{itemize}

We present the limitations and threats to the validity of our dataset as follows, according to \cite{wohlin_experimentation_2024}.

\textit{Internal Validity:} The manual annotation of SATD comments by the two authors may introduce annotator bias, affecting the reliability of the classifications in the CppSATD dataset. We mitigated this bias by adhering to the detailed annotation RD developed in Section \ref{data_annotate}, which contains comprehensive definitions and examples for each SATD type included in the CppSATD dataset.

\textit{Construct Validity:} As we treat the remaining comments after ETF and HTF-pattern filtering as NON-SATD, there is a concern that there are still SATD comments hidden within this remaining set. However, we took a statistically significant sample of 16,125 comments and manually annotated this sample, from which we could conclude at a confidence level of 99\% that the proportion of SATD comments in the remaining set does not exceed 1\%, thereby supporting the assumption that the majority of this remaining set is truly NON-SATD.

\textit{External Validity:} Considering the generalizability of our work, the core limitation within the CppSATD dataset is that the selected repositories are highly popular and well-maintained. Thus, this work might not be able to reflect the SATD characteristics exposed in smaller or less mature projects - which are significantly different in terms of code review practices, development team size, documentation culture, and functionalities. However, this is the first C++ multi-class SATD comments dataset that includes the code context of each comment, aiming to build a foundation for research in this programming language. Future work might build upon CppSATD to cover a broader scope of projects.


\section*{SOURCE CODE AND SCRIPTS}
The Python source code and text patterns used for our data extraction are included in the ZIP folder \textbf{cppsatd.data.extraction.zip} of our replication package at Zenodo \cite{pham_2025_15562944}. We present the versions and specific details of the five selected C++ repositories in Table \ref{tab:repo_metadata}. Moreover, to convert source code to XML representations, we used srcML v1.0.0 tool \cite{collard2013srcml}, whose latest release was in January 2020.

\section*{ACKNOWLEDGEMENTS AND INTERESTS}
This work was done independently of any grants or awards. The article authors have declared no conflicts of interest.

\noindent\textbf{Phuoc Pham:} Methodology, Writing - Original Draft.

\noindent\textbf{Murali Sridharan:} Methodology, Writing - Original Draft.

\noindent\textbf{Matteo Esposito:} Methodology, Writing - Review \& Editing, Supervision.

\noindent\textbf{Valentina Lenarduzzi:} Conceptualization, Validation, Supervision Resources, Writing - Review, Project administration.

\bibliographystyle{IEEEtran}
\bibliography{main}

\begin{thebibliography}{10}
\providecommand{\url}[1]{#1}
\csname url@samestyle\endcsname
\providecommand{\newblock}{\relax}
\providecommand{\bibinfo}[2]{#2}
\providecommand{\BIBentrySTDinterwordspacing}{\spaceskip=0pt\relax}
\providecommand{\BIBentryALTinterwordstretchfactor}{4}
\providecommand{\BIBentryALTinterwordspacing}{\spaceskip=\fontdimen2\font plus
\BIBentryALTinterwordstretchfactor\fontdimen3\font minus \fontdimen4\font\relax}
\providecommand{\BIBforeignlanguage}[2]{{%
\expandafter\ifx\csname l@#1\endcsname\relax
\typeout{** WARNING: IEEEtran.bst: No hyphenation pattern has been}%
\typeout{** loaded for the language `#1'. Using the pattern for}%
\typeout{** the default language instead.}%
\else
\language=\csname l@#1\endcsname
\fi
#2}}
\providecommand{\BIBdecl}{\relax}
\BIBdecl

\bibitem{cunningham1992wycash}
W.~Cunningham, ``The wycash portfolio management system,'' \emph{ACM Sigplan Oops Messenger}, vol.~4, no.~2, pp. 29--30, 1992.

\bibitem{esposito_early_2023}
M.~Esposito, A.~Janes, T.~Kilamo, and V.~Lenarduzzi, ``Early {Career} {Developers}’ {Perceptions} of {Code} {Understandability}. {A} {Study} of {Complexity} {Metrics},'' \emph{A Study of Complexity Metrics}, 2023.

\bibitem{esposito_can_2023}
\BIBentryALTinterwordspacing
M.~Esposito, S.~Moreschini, V.~Lenarduzzi, D.~Hästbacka, and D.~Falessi, ``Can {We} {Trust} the {Default} {Vulnerabilities} {Severity}?'' in \emph{23rd {IEEE} {International} {Working} {Conference} on {Source} {Code} {Analysis} and {Manipulation}, {SCAM} 2023, {Bogotá}, {Colombia}, {October} 2-3, 2023}, L.~Moonen, C.~D. Newman, and A.~Gorla, Eds.\hskip 1em plus 0.5em minus 0.4em\relax IEEE, 2023, pp. 265--270. [Online]. Available: \url{https://doi.org/10.1109/SCAM59687.2023.00037}
\BIBentrySTDinterwordspacing

\bibitem{alves_ontology_technical_debt}
N.~S. Alves, L.~F. Ribeiro, V.~Caires, T.~S. Mendes, and R.~O. Spínola, ``Towards an ontology of terms on technical debt,'' in \emph{2014 Sixth International Workshop on Managing Technical Debt}, 2014, pp. 1--7.

\bibitem{da2017using}
E.~da~Silva~Maldonado, E.~Shihab, and N.~Tsantalis, ``Using natural language processing to automatically detect self-admitted technical debt,'' \emph{IEEE Transactions on Software Engineering}, vol.~43, no.~11, pp. 1044--1062, 2017.

\bibitem{sheikhaei2024empirical}
M.~S. Sheikhaei, Y.~Tian, S.~Wang, and B.~Xu, ``An empirical study on the effectiveness of large language models for satd identification and classification,'' \emph{Empirical Software Engineering}, vol.~29, no.~6, p. 159, 2024.

\bibitem{santos_lstm}
R.~M. Santos, M.~C.~R. Junior, and M.~G. de~Mendon{\c{c}}a~Neto, ``Self-admitted technical debt classification using lstm neural network,'' in \emph{17th International Conference on Information Technology--New Generations (ITNG 2020)}, S.~Latifi, Ed.\hskip 1em plus 0.5em minus 0.4em\relax Cham: Springer International Publishing, 2020, pp. 679--685.

\bibitem{chen_xgboost}
X.~Chen, D.~Yu, X.~Fan, L.~Wang, and J.~Chen, ``Multiclass classification for self-admitted technical debt based on xgboost,'' \emph{IEEE Transactions on Reliability}, vol.~71, no.~3, pp. 1309--1324, 2022.

\bibitem{yu_gan}
\BIBentryALTinterwordspacing
J.~Yu, X.~Zhou, X.~Liu, J.~Liu, Z.~Xie, and K.~Zhao, ``Detecting multi-type self-admitted technical debt with generative adversarial network-based neural networks,'' \emph{Information and Software Technology}, vol. 158, p. 107190, 2023. [Online]. Available: \url{https://www.sciencedirect.com/science/article/pii/S0950584923000447}
\BIBentrySTDinterwordspacing

\bibitem{farias_code_comment_analysis}
\BIBentryALTinterwordspacing
M.~A. de~Freitas~Farias, M.~G. de~Mendonça~Neto, M.~Kalinowski, and R.~O. Spínola, ``Identifying self-admitted technical debt through code comment analysis with a contextualized vocabulary,'' \emph{Information and Software Technology}, vol. 121, p. 106270, 2020. [Online]. Available: \url{https://www.sciencedirect.com/science/article/pii/S0950584920300203}
\BIBentrySTDinterwordspacing

\bibitem{model2011notation}
B.~P. Model, ``Notation (bpmn) version 2.0,'' \emph{OMG Specification, Object Management Group}, vol.~19, pp. 52--60, 2011.

\bibitem{sridharan2023pentacet}
M.~Sridharan, L.~Rantala, and M.~M{\"a}ntyl{\"a}, ``Pentacet data-23 million contextual code comments and 250,000 satd comments,'' in \emph{2023 IEEE/ACM 20th International Conference on Mining Software Repositories (MSR)}.\hskip 1em plus 0.5em minus 0.4em\relax IEEE, 2023, pp. 412--416.

\bibitem{rubaye_popularity}
A.~Al-Rubaye and G.~Sukthankar, ``Scoring popularity in github,'' in \emph{2020 International Conference on Computational Science and Computational Intelligence (CSCI)}, 2020, pp. 217--223.

\bibitem{collard2013srcml}
M.~L. Collard, M.~J. Decker, and J.~I. Maletic, ``srcml: An infrastructure for the exploration, analysis, and manipulation of source code: A tool demonstration,'' in \emph{2013 IEEE International conference on software maintenance}.\hskip 1em plus 0.5em minus 0.4em\relax IEEE, 2013, pp. 516--519.

\bibitem{maldonado_2015}
E.~d.~S. Maldonado and E.~Shihab, ``Detecting and quantifying different types of self-admitted technical debt,'' in \emph{2015 IEEE 7th International Workshop on Managing Technical Debt (MTD)}, 2015, pp. 9--15.

\bibitem{cohen1960coefficient}
J.~Cohen, ``A coefficient of agreement for nominal scales,'' \emph{Educational and psychological measurement}, vol.~20, no.~1, pp. 37--46, 1960.

\bibitem{falessi_enhancing_2023}
\BIBentryALTinterwordspacing
D.~Falessi, S.~M. Laureani, J.~Çarka, M.~Esposito, and D.~A.~d. Costa, ``Enhancing the defectiveness prediction of methods and classes via {JIT},'' \emph{Empir. Softw. Eng.}, vol.~28, no.~2, p.~37, 2023. [Online]. Available: \url{https://doi.org/10.1007/s10664-022-10261-z}
\BIBentrySTDinterwordspacing

\bibitem{esposito_extensive_2024}
\BIBentryALTinterwordspacing
M.~Esposito, V.~Falaschi, and D.~Falessi, ``An {Extensive} {Comparison} of {Static} {Application} {Security} {Testing} {Tools},'' in \emph{Proceedings of the 28th {International} {Conference} on {Evaluation} and {Assessment} in {Software} {Engineering}, {EASE} 2024, {Salerno}, {Italy}, {June} 18-21, 2024}.\hskip 1em plus 0.5em minus 0.4em\relax ACM, 2024, pp. 69--78. [Online]. Available: \url{https://doi.org/10.1145/3661167.3661199}
\BIBentrySTDinterwordspacing

\bibitem{SimWright2005}
J.~Sim and C.~C. Wright, ``The kappa statistic in reliability studies: Use, interpretation, and sample size requirements,'' \emph{Physical Therapy}, vol.~85, no.~3, pp. 257--268, 03 2005.

\bibitem{esposito_beyond_2024}
\BIBentryALTinterwordspacing
M.~Esposito, F.~Palagiano, V.~Lenarduzzi, and D.~Taibi, ``Beyond {Words}: {On} {Large} {Language} {Models} {Actionability} in {Mission}-{Critical} {Risk} {Analysis},'' in \emph{Proceedings of the 18th {ACM}/{IEEE} {International} {Symposium} on {Empirical} {Software} {Engineering} and {Measurement}, {ESEM} 2024, {Barcelona}, {Spain}, {October} 24-25, 2024}, X.~Franch, M.~Daneva, S.~Martínez-Fernández, and L.~Quaranta, Eds.\hskip 1em plus 0.5em minus 0.4em\relax ACM, 2024, pp. 517--527. [Online]. Available: \url{https://doi.org/10.1145/3674805.3695401}
\BIBentrySTDinterwordspacing

\bibitem{esposito_leveraging_2024}
\BIBentryALTinterwordspacing
M.~Esposito and F.~Palagiano, ``Leveraging {Large} {Language} {Models} for {Preliminary} {Security} {Risk} {Analysis}: {A} {Mission}-{Critical} {Case} {Study},'' in \emph{Proceedings of the 28th {International} {Conference} on {Evaluation} and {Assessment} in {Software} {Engineering}, {EASE} 2024, {Salerno}, {Italy}, {June} 18-21, 2024}.\hskip 1em plus 0.5em minus 0.4em\relax ACM, 2024, pp. 442--445. [Online]. Available: \url{https://doi.org/10.1145/3661167.3661226}
\BIBentrySTDinterwordspacing

\bibitem{pham_2025_15562944}
\BIBentryALTinterwordspacing
P.~Pham, M.~Sridharan, M.~Esposito, and V.~Lenarduzzi, ``Cppsatd: A reusable self-admitted technical debt dataset in c++,'' May 2025. [Online]. Available: \url{https://doi.org/10.5281/zenodo.15562944}
\BIBentrySTDinterwordspacing

\bibitem{potdar2014exploratory}
A.~Potdar and E.~Shihab, ``An exploratory study on self-admitted technical debt,'' in \emph{2014 IEEE International Conference on Software Maintenance and Evolution}.\hskip 1em plus 0.5em minus 0.4em\relax IEEE, 2014, pp. 91--100.

\bibitem{digkas_evolution}
G.~Digkas, M.~Lungu, A.~Chatzigeorgiou, and P.~Avgeriou, ``The evolution of technical debt in the apache ecosystem,'' in \emph{Software Architecture: 11th European Conference, ECSA 2017, Canterbury, UK, September 11-15, 2017, Proceedings 11}.\hskip 1em plus 0.5em minus 0.4em\relax Springer, 2017, pp. 51--66.

\bibitem{sutoyo_tracing}
\BIBentryALTinterwordspacing
E.~Sutoyo, P.~Avgeriou, and A.~Capiluppi, ``Tracing the lifecycle of architecture technical debt in software systems: A dependency approach,'' 2025. [Online]. Available: \url{https://arxiv.org/abs/2501.15387}
\BIBentrySTDinterwordspacing

\bibitem{esposito_generative_2025}
M.~Esposito, X.~Li, S.~Moreschini, N.~Ahmad, T.~Cerny, K.~Vaidhyanathan, V.~Lenarduzzi, and D.~Taibi, ``Generative {AI} for {Software} {Architecture}. {Applications}, {Trends}, {Challenges}, and {Future} {Directions},'' \emph{arXiv preprint arXiv:2503.13310}, 2025.

\bibitem{wohlin_experimentation_2024}
\BIBentryALTinterwordspacing
C.~Wohlin, P.~Runeson, M.~Höst, M.~C. Ohlsson, B.~Regnell, and A.~Wesslén, \emph{Experimentation in {Software} {Engineering}, {Second} {Edition}}.\hskip 1em plus 0.5em minus 0.4em\relax Springer, 2024. [Online]. Available: \url{https://doi.org/10.1007/978-3-662-69306-3}
\BIBentrySTDinterwordspacing

\end{thebibliography}

\end{document}